\documentclass[conference]{IEEEtran}
\IEEEoverridecommandlockouts%
\usepackage{url}
\usepackage{cite}
\usepackage{booktabs}
\usepackage{multirow}
\usepackage{enumerate}

\usepackage{amsmath,amssymb,amsfonts}
\usepackage{textcomp}
\usepackage{balance}

\usepackage{todonotes}
\presetkeys{todonotes}{inline}{}

\usepackage{tikz}
\usetikzlibrary{arrows,arrows.meta,backgrounds,calc,decorations.pathreplacing,decorations.markings,decorations.pathmorphing,decorations.pathreplacing,external,colorbrewer,intersections,shadows,shapes,shapes.misc}
\tikzexternaldisable%

\usepackage{pgfplots}
\usepgfplotslibrary{groupplots,colormaps,colorbrewer,statistics}
\usepackage{pgfplotstable}
\pgfplotsset{compat=1.15}

\usepackage{pifont}
\newcommand{\cmark}{\ding{51}}%

\usepackage{xspace}
\newcommand{\eg}{\textit{e.g.,}~}
\newcommand{\ie}{\textit{i.e.,}~}

\newcommand{\etal}{\textit{et al.}~}
\newcommand{\one}{({\em i})\xspace}
\newcommand{\two}{({\em ii})\xspace}
\newcommand{\three}{({\em iii})\xspace}

\makeatletter
\renewcommand{\paragraph}[1]{\vspace*{0.03in}\noindent{\bf #1.}\hspace{0.25ex \@plus1ex \@minus.2ex}}
\newcommand{\paragraphb}[1]{\vspace*{0.03in}\noindent{\bf #1}\hspace{0.25ex \@plus1ex \@minus.2ex}}
\makeatother

\begin{document}

\title{Networking Group Content: RESTful Multiparty Access to a Data-centric Web of Things}

\author{\IEEEauthorblockN{Cenk G\"undo\u{g}an}
\IEEEauthorblockA{\textit{HAW Hamburg}\\
\small cenk.guendogan@haw-hamburg.de}
\and
\IEEEauthorblockN{Christian Ams\"uss}
\IEEEauthorblockA{\textit{~} \\
\small christian@amsuess.com}
\and
\IEEEauthorblockN{Thomas C. Schmidt}
\IEEEauthorblockA{\textit{HAW Hamburg} \\
\small t.schmidt@haw-hamburg.de}
\and
\IEEEauthorblockN{Matthias W\"ahlisch}
\IEEEauthorblockA{\textit{Freie Universit\"at Berlin} \\
\small m.waehlisch@fu-berlin.de}
}

\maketitle

\begin{abstract}
Content replication to many destinations is a common use case in the Internet of Things (IoT). The deployment of IP multicast has proven inefficient, though, due to its lack of layer-2 support by common IoT radio technologies and its synchronous end-to-end transmission, which is highly susceptible to interference. Information-centric networking (ICN) introduced hop-wise multi-party dissemination of cacheable content, which  has  proven valuable in particular for low-power lossy networking regimes. Even NDN, however, the most prominent ICN protocol, suffers from a lack of deployment.
 
In this paper, we explore how multiparty content distribution in an information-centric Web of Things (WoT) can be built on CoAP\@. We augment the CoAP proxy by request aggregation and response replication functions, which together with proxy caches enable asynchronous group communication. In a further step, we integrate content object security with OSCORE into the CoAP multicast proxy system, which enables ubiquitous caching of certified authentic content. 	
In our evaluation, we compare NDN with different deployment models of CoAP, including our data-centric approach in realistic testbed experiments. Our findings indicate that multiparty content distribution based on CoAP proxies performs equally well as NDN, while remaining fully compatible with the established IoT protocol world of CoAP on the Internet.     
\end{abstract}


\section{Introduction}\label{sec:intro}

The Internet of Things (IoT) is about to interconnect billions of embedded controllers via the Internet using protocols that adapt to its constrained regime. The wireless edge deploys 6LoWPAN~\cite{RFC-4944}  as an adaptation layer of IPv6 to low-power lossy links and applications exchange content via CoAP~\cite{RFC-7252} as a lightweight alternative to HTTP\@. It has been shown many times, though, that a plain translation of Internet protocols turns inefficient in the data-centric Web of Things. Instead, information-centric networking (ICN) approaches~\cite{adiko-sind-12} proved significantly more robust and reliable by providing hop-wise forwarding with content caching in the lossy wireless regimes.

One challenging example is multiparty communication. While the information-centric named-data networking (NDN)~\cite{jstp-nnc-09} architecture supports seamless replication of group content, IP multicast is difficult to implement in low-power wireless regimes due to a lack of layer 2 support. In addition, the synchronous nature of IP multicast endangers successful packet transmissions due to interferences. Multiparty content dissemination, however, is an important use case in common actuator scenarios (\eg switching light bulbs) and in particular for over the air software updates.

In this paper, we construct a data-centric Web of Things (WoT) that can aggregate content requests and replicate  responses using CoAP protocol elements.
We extend existing CoAP components with few additional functions, which we implemented in the RIOT IoT operating system. These extended CoAP components provide all required ICN-type properties: hop-wise multi-destination forwarding, on-path caching, request aggregation, response fan-out, and group access to secured content objects. We evaluate our data-centric WoT model against NDN and plain CoAP variants. Our findings indicate that our approach performs similar to NDN, while largely outperforming plain CoAP deployments.

The remainder of this paper is structured as follows. We introduce the problem and related work in Section~\ref{sec:relatedwork}. Section~\ref{sec:multiparty} presents our protocol approach along with a discussion of its information-centric properties. We evaluate comparatively evaluate our model against NDN and different CoAP configurations in Section~\ref{sec:evaluation}, and finally conclude with an outlook in Section~\ref{sec:conclusion}.


\section{The Problem of Multicast in the IoT\\and Related Work}\label{sec:relatedwork}
\begin{figure}
  \centering
  \includegraphics{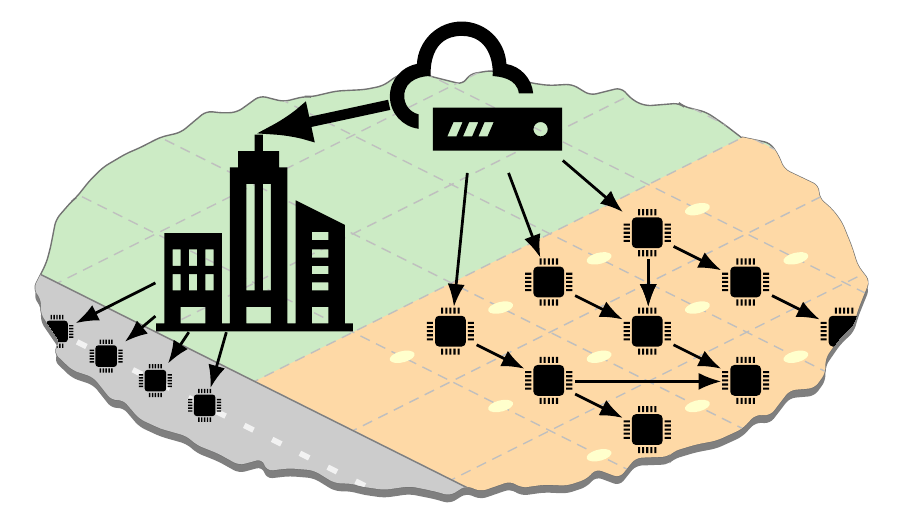}
  \caption{Massive firmware roll-outs in distributed and heterogeneous networks.}%
  \label{fig:overview}
\end{figure}

\begin{figure*}
  \centering
  \includegraphics{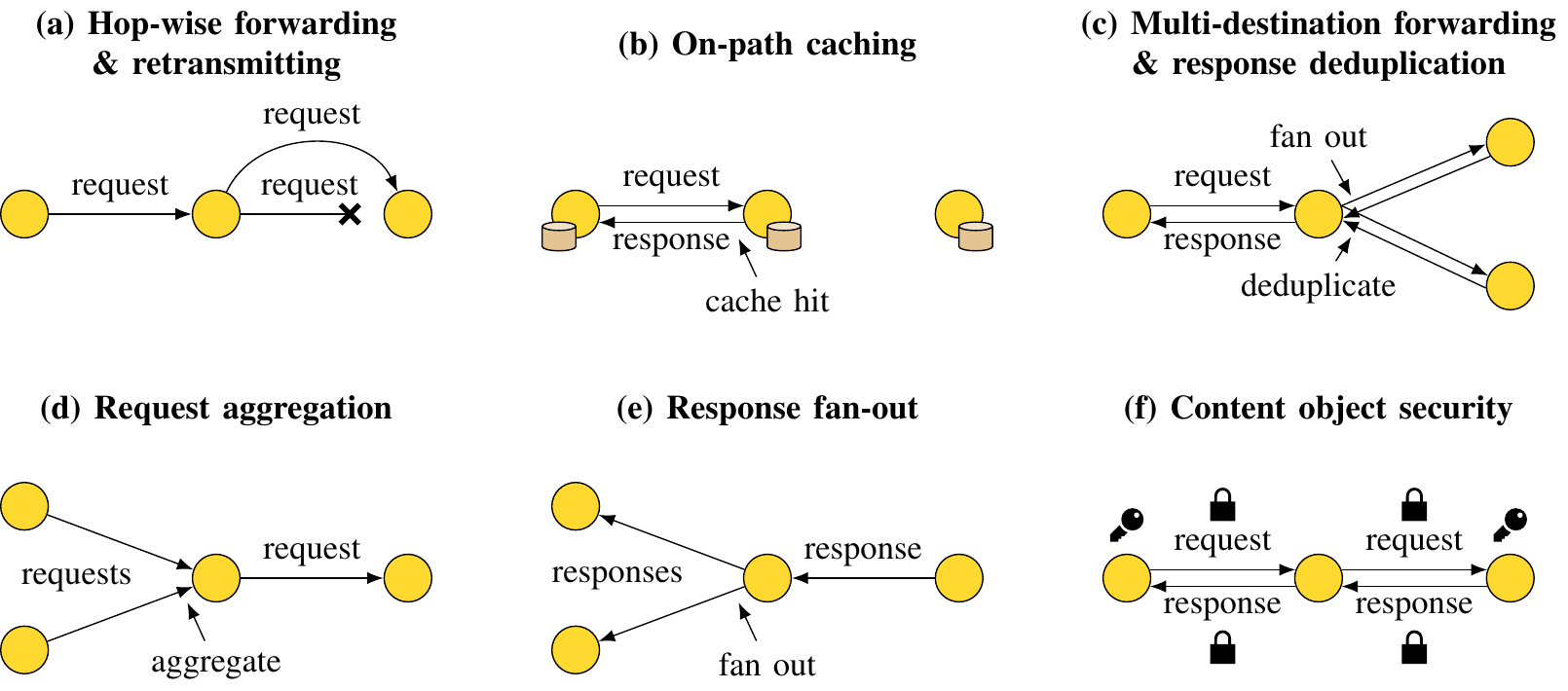}
  \caption{NDN protocol features that enable an efficient and secure multiparty communication for the IoT.}%
  \label{fig:ndn-features}
\end{figure*}

\subsection{Challenges of IoT Group Scenarios}
IoT message exchange follows the patterns `scheduled' and `on demand' between individual node pairs or in groups. Group communication is desired for data fusion, \eg when a group of sensors returns its readings following a single subscription or data request. Group communication is also needed for disseminating data to large sets of receivers, \eg for distributing instructions to actuators, and may consist of large data volumes, for instance in the case of software updates (see \figurename~\ref{fig:overview}).  

Group communication requires scalable network solutions, whenever an iterated unicast transmission between participants will impose critical stress onto network links. Such limits are quickly reached in low-power lossy wireless regimes for which rates of 100 packets/s may already strain a link.
With battery-operated or energy-harvesting devices, energy conditions are often even more critical. Traffic flows greatly dominate energy expenditures~\cite{lpcs-tsacp-04} and hence the lifespan of the involved nodes.

\subsection{Multicast and Its Limitations in the IoT}
The traditional approach to group communication on the Internet is IP multicast~\cite{RFC-1112}. Multicast network costs scale well as a root of the number of receivers~\cite{mhh-oem-01}, and IP multicast seamlessly hosts stateless UDP transport, which dominates the IoT. Further, mappings exist to common link layers of local area networks such as Ethernet or WLAN\@.

Constrained wireless technologies, however, that do not employ any form of slotted channel access, but rather use carrier-sensing (\eg IEEE~802.15.4~\cite{IEEE-802.15.4-16}) or pure ALOHA (\eg LoRAWAN~\cite{lorawan-spec-11}), typically lack support for multicast on the link-layer and default to broadcast.
Also for Bluetooth Low Energy (BLE)~\cite{b-bcsv-19}, which utilizes proper multiplexing schemes by the frequency and time domain to reduce radio listening cycles, an efficient IP multicast mapping is not given.
Since BLE connections between devices are point-to-point, an IP multicast is realized by duplicating multicast messages on each unicast link~\cite[Section 3.2.5]{RFC-7668}.

Multicast routing in the Internet is complex and multicast mobility adds further complications~\cite{RFC-5757}.
Bit Index Explicit Replication (BIER)~\cite{RFC-8279} is a new multicast architecture that promises a considerable simplification by eliminating per-flow state from  routers.
The Multicast Protocol for Low-Power and Lossy Networks (MPL)~\cite{RFC-7731} establishes IP multicast forwarding in constrained and wireless deployments.
Instead of building a dissemination tree, MPL uses a controlled flooding approach combined with a mechanism to detect and suppress the propagation of duplicate messages.
Other approaches~\cite{draft-ietf-roll-ccast-01,draft-thubert-roll-bier-02} supplement the IPv6 Routing Protocol for Low-Power and Lossy Networks (RPL)~\cite{RFC-6550} to operate in a similar fashion as BIER\@.

CoAP extensions~\cite{RFC-7390,draft-ietf-core-groupcomm-bis-02,draft-tiloca-core-groupcomm-proxy-02} update the request-response model to enable a one-to-many communication using multicast IP addresses for resource endpoints on the application layer.
While this amendment allows clients to perform requests to a group of CoAP servers, the resulting responses always return as unicast to the respective client.
To remain stateless, a CoAP group communication only allows non-confirmable multicast requests~\cite[Section 2.3.1]{draft-ietf-core-groupcomm-bis-02}.

The protection of communication flows between multiple endpoints poses another challenge.
Transport layer security is the default strategy to deliver protective measures for streams in the Internet~\cite{RFC-8446} and for datagrams in the IoT~\cite{RFC-6347} between two hosts.
Security contexts are tightly bound to socket endpoints and use the five-tuple (IP$_{src}$, Port$_{src}$, IP$_{dst}$, Port$_{dst}$, Transport) to identify secured channels.
This strong binding makes transport layer security impractical in flows involving multiple devices.


\section{Multiparty Content Retrieval for a Data-centric Web of Things}\label{sec:multiparty}

\subsection{Named-Data Networking for the IoT}

Named-Data Networking (NDN)~\cite{jstp-nnc-09} is a future Internet architecture that follows the information-centric networking (ICN) paradigm.
NDN employs a stateful and pull-driven forwarding fabric with built-in reliability features and multicast support, which focuses around the request-response communication pattern.
In comparison to the general-purpose Internet, where only a single IP datagram exists to encapsulate upper layer packets, NDN specifies two message types on the network layer with contrasting semantics:~\textit{Interests} for requesting content and \textit{Data} for delivering responses.
NDN exhibits protocol features (see \figurename~\ref{fig:ndn-features}) that have been proven valuable in constrained and wireless IoT networks~\cite{sblwy-ndnti-16,mwt-tucin-16,gklp-ncmcm-18}.

\subsection{NDN Protocol Features}\label{subsec:ndn-protocol-features}

\paragraph{Hop-wise Forwarding \& Retransmitting}
Interest messages traverse hop-by-hop and are forwarded on human-readable names akin to URLs for web resources.
Each forwarder tracks state for open requests in the \textit{Pending Interest Table}.
Data messages return on the constructed request path and consume the existing forwarding state.
When a response is lost as illustrated in \figurename~\ref{fig:ndn-features} (a), hop-wise timeouts occur and requests are retransmitted.
Retransmissions are confined to links that have not been traversed by responses yet.

\paragraph{On-path Caching}
Individual hops maintain a content store as an integral part of the forwarding logic.
Data are stored in this cache and returned for matching requests as displayed in \figurename~\ref{fig:ndn-features} (b).
Caching provides location-independence for content and is a fundamental feature to improve bandwidth and latency of content retrievals.
Especially in low-power setups with intermittent connectivity, caching paired with the previously discussed corrective behavior yields an increased robustness due to shortened request paths on retransmissions.

\paragraph{Multi-destination Forwarding \& Response Deduplication}
The Forwarding Information Base (FIB) records content names and outgoing faces.
One compelling difference to an IP FIB is that multiple faces can exist for a single destination to enable one-to-many flows (see \figurename~\ref{fig:ndn-features} (c)).
A forwarding strategy commits to either all outgoing faces at once, or makes sensible decisions to select a subset.
If multiple responses return as a result of request fan-outs, then they are deduplicated.

\paragraph{Request Aggregation \& Response Fan-out}
Simultaneous requests from multiple origins can meet on shared paths as shown in \figurename~\ref{fig:ndn-features} (d).
Messages aggregate on hops that previously set up appropriate forwarding state.
Incoming faces of equal requests are cataloged and when a response returns, it fans out to all stored faces (see \figurename~\ref{fig:ndn-features} (e)).

\paragraph{Content Object Security}
Unlike schemes that focus on transport layer security between endpoints, messages are cryptographically signed and content can be encrypted without requiring any endpoint information (see \figurename~\ref{fig:ndn-features} (f)).
Content object security allows content caching for long periods of network disruption, while preserving all security measures without maintaining endpoint-based security contexts.

\subsection{An Information-centric Web of Things}\label{subsec:icnwot}
In previous research, G\"undo\u{g}an \etal\cite{gasw-triwt-20} designed an information-centric Web of Things (WoT) that uses CoAP forward proxies on every hop along a path.
This deployment embodies a subset of the NDN protocol features displayed in Section~\ref{subsec:ndn-protocol-features}.
As their comparative measurements show, a hop-wise forwarding, hop-by-hop retransmissions, and response caching on each proxy node are sufficient to boost the network resilience up to the level of standard NDN setups.

However, their architectural design misses these integral features to enable an efficient multiparty communication: \one multi-destination forwarding with response deduplication, \two request aggregation from multiple origins with response fan-out, and \three a pluralistic cache utilization.
Their na\"ive OSCORE~\cite{RFC-8613} integration provides content object security, but confines the effects of caching and request aggregation only to request-response pairs due to the strong message binding that OSCORE introduces.
With this limitation, only retransmissions benefit from information-centric properties.

In the remainder, we will extend the information-centric WoT construction with support for multiparty content access by integrating the missing features into the CoAP deployment.

\subsection{Multi-destination Forwarding}\label{subsec:multidest-forwarding}
\begin{figure}[h]
  \centering
  \includegraphics{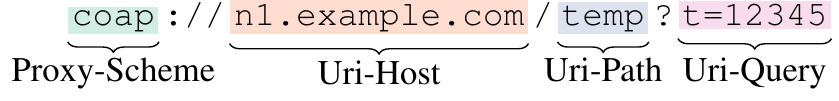}
  \caption{Equivalent components of a Proxy-Uri as used with forward proxies.}\label{fig:proxy-uri}
\end{figure}

When a CoAP node attempts to obtain a resource representation, it encodes the resource URI in a set of CoAP options, either in the \textit{Proxy-Uri} option or using individual options (see \figurename~\ref{fig:proxy-uri}).
Note that while the orange part is named \textit{host} due to its predominant use,
it may use any locally defined lookup system~\cite[Section 3.2.2]{RFC-3986}.

Nodes may send a request to the known network address of the server,
or to a usually more powerful proxy node to defer routing, name resolution, and protocol conversion complexity.
In the common case of deferred routing, the proxy will use DNS and its routing table to send the request to the server.

In the information-centric WoT design (see Section~\ref{subsec:icnwot}) forward proxies are used on each hop along a path, and the Proxy-Uri option is present as a forwarding hint in all hop-wise requests but the last.
On that final hop, the scheme and host information can be discarded, and the more compact Uri-Path and Uri-Query are sent instead of Proxy-Uri.

\begin{table}[h]
  \centering
  \begin{tabular*}{\columnwidth}{l @{\extracolsep{\fill}} l c}
    \toprule
    \textbf{URI pattern} & \textbf{Next-hop} & \textbf{Send host}\\
    \midrule
    \multirow{2}{*}{coap://00-01/temperature*} & coap://[fe80::1\%0] & yes \\
    & coap://[fe80::2\%1] & yes \\[0.40em]
    {coap://00-02/firmware/*} & coap://[fe80::3\%0] & no \\
    \bottomrule\\
  \end{tabular*}
  \caption{Application-level Forwarding Information Base for CoAP.}\label{tab:icnwot-fib}
\end{table}

A FIB structure (see \tablename~\ref{tab:icnwot-fib}) managed on the proxy stores the next-hop, and whether that hop requires a full Proxy-Uri option including the \textit{host} component.
Analogous to NDN, we adjust this FIB to allow multiple next-hop entries for a single destination.
If multiple next-hops exist, the request is duplicated to all available endpoints.
As CoAP requests are not necessarily idempotent (some are not even side-effect free),
requests with codes like POST or PATCH still have to take a single next-hop.

\subsection{Response Deduplication}
Duplicate responses return to a forwarder node if requests are replicated onto different paths as part of the multi-destination forwarding process (see Section~\ref{subsec:multidest-forwarding}).
A forwarder has two choices when tasked with the deduplication of response messages.
First, if content is dynamic and not bound to the resource path, then all returning responses are aggregated to form a single representative.
An in-network deduplication, however, has the issues of deducing correct delays to capture all returning messages and to apply a reduce function of arbitrary complexity depending on the use case.
The second choice consists of forwarding only the first arriving response message, while discarding all other occurrences.
This case is more efficient, because it reduces link stress to a minimum, but is only effective in deployments where content binds immutably to resource paths.
In our design considerations for an information-centric architecture, we opt for the second choice as it is conforming with the philosophy of NDN to have a strong name to content binding.

In CoAP, responses match to corresponding requests with the use of token values.
A client generates a token and includes it in a request.
To make the required CoAP state for requests similarly compact as in NDN,
clients can use a shared token space for all their next-hops.
Then, incoming responses consume that request state,
and get forwarded to all interested clients.
Any further responses based on that token
are rejected or ignored as illustrated in \figurename~\ref{fig:ndn-features} (c).

\subsection{Request Aggregation and Response Fan-out}

\begin{figure}[h]
  \centering
  \includegraphics{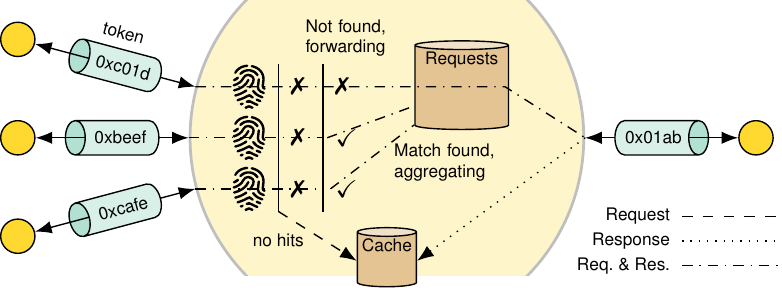}
  \caption{Forwarding logic for request aggregation and response fan-out.}\label{fig:reqaggfanout}
\end{figure}

Simultaneous requests to the same resource path from different clients can be aggregated to reduce link stress.
In parallel to checking for cached responses using the CoAP Cache-Key~\cite[Section 5.6]{RFC-7252} based on CoAP options, the set of open request states is checked using the same key.
If state exists, then a forwarder records the newly arriving request along with the existing information.
Tokens in client requests provide a unique identification for transactions and must be stored individually to preserve the request-response matching as illustrated in \figurename~\ref{fig:reqaggfanout}.

For the very first request, the forwarder sends out a new request message that is semantically equivalent to the original request.
Tokens (along with some other message properties) are local to the hop, and thus generally differ between incoming and forwareded requests.
When a response returns, the list of interested clients contains their stored addresses and tokens,
from which corresponding responses are built and fanned out.

\subsection{The Problem with Content Object Security}

\begin{figure}[h]
  \centering
  \includegraphics{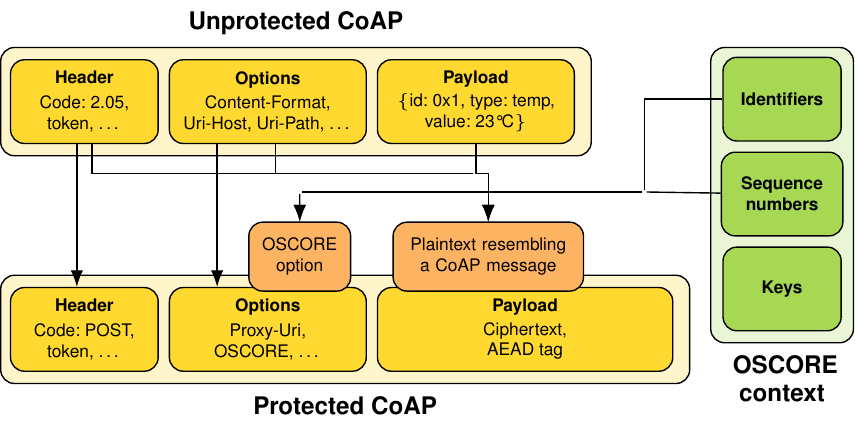}
  \caption{COSE and OSCORE to protect CoAP messages on the object level.}\label{fig:oscore}
\end{figure}

The information-centric WoT deployment leverages OSCORE~\cite{RFC-8613} to protect CoAP messages across network boundaries.
It provides request and response confidentiality,
integrity across request and response,
and source authentication even in its group mode~\cite{draft-ietf-core-oscore-groupcomm}.
A COSE~\cite{RFC-8152} object is populated with a statically preconfigured or dynamically derived~\cite{draft-ietf-lake-edhoc} OSCORE context.
While a COSE header holds meta information, \eg key identifiers and encryption algorithm, a COSE ciphertext contains parts of a CoAP message that are considered for encryption.
A protected CoAP message then carries the COSE header as an OSCORE option and the COSE ciphertext as payload (see \figurename~\ref{fig:oscore}).

OSCORE complements the information-centric WoT with security on the content object level to enable secured CoAP deployments that are on par with NDN setups in terms of network performance.
There are, however, two aspects that greatly affect the carefully designed multiparty content access functionality when OSCORE is na\"ively employed on request paths that solely consist of proxy nodes: \one reduced cache utilization and \two impaired routing decisions.

\paragraph{Caching}
Clients generate unique nonces that are used by cryptographic operations for request-response pairs.
With the nonce both directly (in the OSCORE option) and indirectly (in the ciphertext) altering the Cache-Key,
distinct OSCORE requests can not utilize caches.
Even if cache hits could produce an equivalent earlier response,
the request-response binding would fail to perform the authenticated encryption with associated data (AEAD) decryption.
The only exception where caching OSCORE messages shows effective results is for request retransmissions on packet loss~\cite{gasw-triwt-20}.

Our architectural decision for re-activating the caching support is to use deterministic requests~\cite{draft-amsuess-core-cachable-oscore}.
It uses OSCORE group communication and introduces the deterministic client,
a fictitious group member which avoids nonce reuse not by using sequence numbers in nonces,
but by hashing the request plaintext and additional data into the key.
Any member of the group can assume the role of the deterministic client by asking the group manager for the correct key details.
Requests with identical plaintext sent from any group member results in the same hash
and identical ciphertext.
These requests carry the full hash in a dedicated CoAP option, which is also included in the calculation of the Cache-Key.
A server receiving such a request uses the sent hash to derive the cryptographic keys,
decrypts, verifies the hash from the plain text,
and responds in group mode for any group member to use the response.

Deterministic OSCORE weakens three properties of OSCORE\@:
\one the request-response binding (responses are only bound to \textit{a} request of the same content, not \textit{the} request the client created),
\two source authentication for requests (which is tolerable as the mechanism only applies to side-effect free requests),
and \three request confidentiality (but only to the extent that an adversary can see that two requests are equal).
Consequently, deterministic requests bypass the replay protections of OSCORE, which is not a significant issue if only idempotent requests are used.
Deterministic OSCORE preserves source authentication for responses by using asymmetric cryptographic signatures.

\paragraph{Routing Decisions with Request Confidentiality}
OSCORE prescribes~\cite[Section~4.1.3.]{RFC-8613} a different treatment for options that form the request URI (see \figurename~\ref{fig:proxy-uri}):
while path and query parameters are encrypted and thus privacy protected,
scheme and host are neither encrypted nor integrity protected.
For the latter, an implicit protection is in place:
The cryptographic context choice of the client ensures that only the right server can respond.
This reflects the HTTP practice that any request to the same origin (\ie scheme, host and port)
is protected in a single connection.
It also enables a flexible application layer routing using forward or reverse proxies.
For name-based FIBs like in NDN, this severly limits the information available for forwarding decisions unless the forwarder is a group member.
Information can still be used in hash based path choices (\eg to load balance across several larger caches)
but not by expectations of response characteristics like in \tablename~\ref{tab:icnwot-fib}.

As the proposed WoT construction does not depend on the host component to map directly to a host,
there is room for application designers to move information between the host and path components.
They can put more information into the unencrypted host component (in the \tablename~\ref{tab:icnwot-fib} example, use coap://00-02-firmware/ instead of coap://00-02/firmware/)
and thus expose more request information,
or move more information into the path component (\eg coap://example.com/00-02/firmware/)
and thus hide more information at the cost of less configurable routing.
In this extreme form, nodes are practically required to be group members to make meaningful routing decisions.


\section{Experimental Evaluation}\label{sec:evaluation}

We experimentally assess the protocol performance of the information-centric WoT in a multiparty content retrieval scenario and compare it against NDN using real IoT hardware.

\subsection{Experiment Setup}

\begin{figure}
  \centering
  \includegraphics{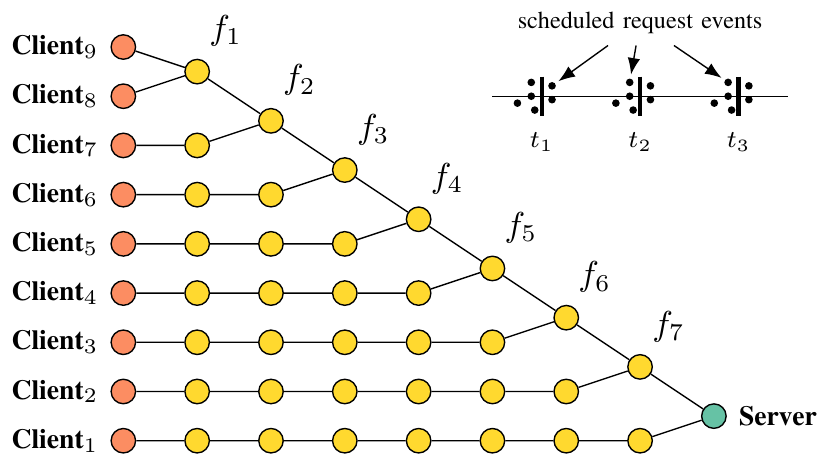}%
  \caption{Testbed topology modeling fan-outs from rank zero to seven of the forwarding hierarchy.}%
  \label{fig:topology}
\end{figure}

\subsubsection{Scenario}
Our experiments implement a typical comand-and-control scenario where multiple actuators show interest in up-to-date instructions from a control node.
For our evaluations, we construct the network topology as depicted in \figurename~\ref{fig:topology}.
The gradual addition of intersections and the long path stretch allow for observing protocol behaviors and inspecting the performances in a better nuanced event space.
A set of nine clients connects to a server that periodically assesses sensory data and the environmental situation to generate timely instructions every second.
In turn, all clients request the latest instruction by appending an increasing sequence number as time offset to the resource name: \texttt{/instruction?t=$x$}.
At experiment begin, we synchronize each client to start the scheduled request pattern simultaneously and clients apply a random jitter in the hundreds of milliseconds range to each request.
In total, each client triggers 1000 requests.

The path between client$_9$ and server demonstrates the most intersections and it is reasonable to assume the highest traffic load for this route.
All forwarders on this path are named $f_{i}$ for easier reference in the evaluation.

\subsubsection{Hardware and Software Platform}
We conduct our experiments on the \texttt{grenoble} site of the IoT-LAB testbed~\cite{abfhm-filso-15}.
It provides a large multi-hop deployment consisting of various class~2~\cite{RFC-7228} devices featuring a 32-bit ARM Cortex-M3 $\mu$controller with 64~kB of RAM and 512~kB of ROM\@.
The platform further mounts an Atmel AT86RF231~\cite{a-lptzi-09} transceiver to operate on the 2.4~GHz IEEE~802.15.4 radio.
Devices run the RIOT~\cite{bghkl-rosos-18} operating system in version 2021.01.
OSCORE experiments use \texttt{libOSCORE}\footnote{https://gitlab.com/oscore/liboscore} and the modular GNRC IPv6 network stack, while the NDN setup uses CCN-lite~\cite{ccn-lite}.

\begin{figure*}
  \centering
  \includegraphics{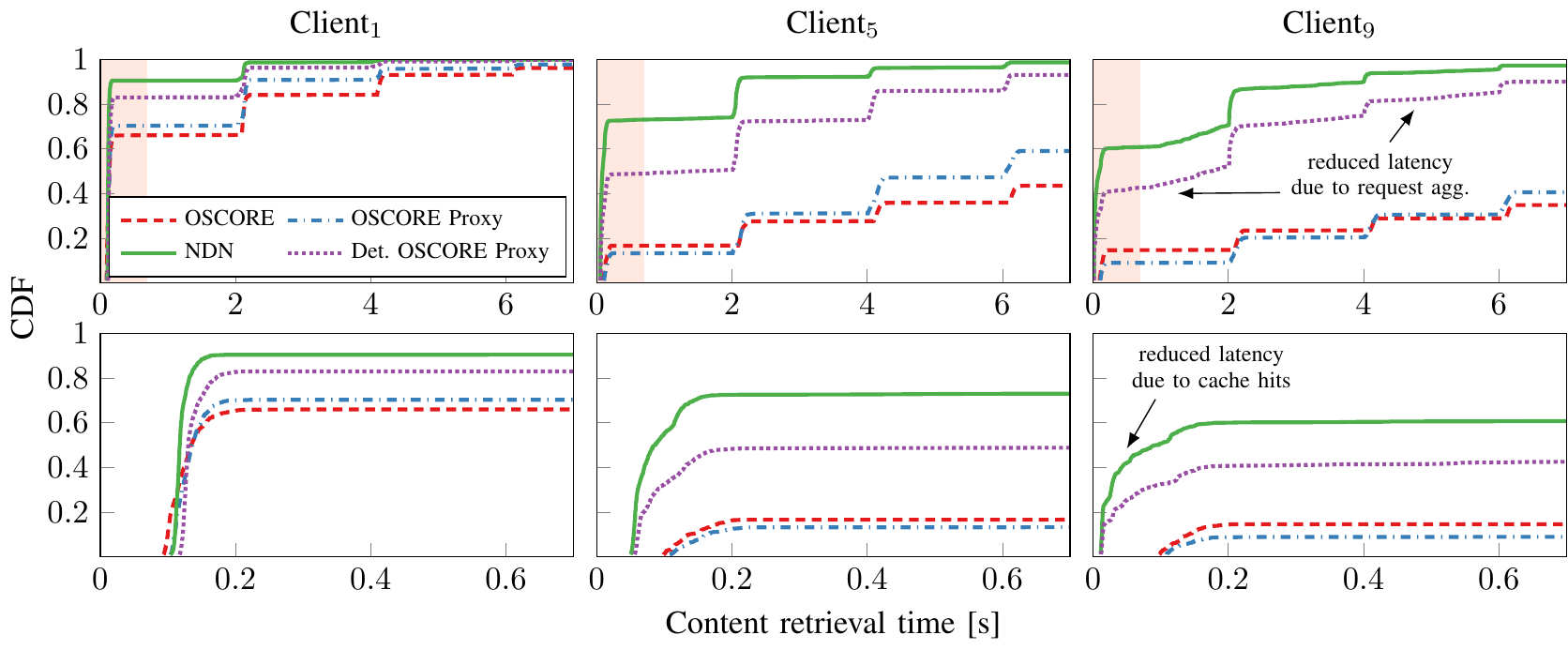}%
  \caption{Content retrieval times---distributions for different client devices and protocol deployments.}%
  \label{fig:completiontime-cdf}
\end{figure*}
\subsubsection{Protocol Settings}
We compare four different protocol deployments with varying degrees of multiparty support and summarize their features in \tablename~\ref{tab:protocol-settings}.
\begin{enumerate}[i)]
\item OSCORE\@: end-to-end CoAP deployment that uses OSCORE to protect single request-response transactions.
\item OSCORE Proxy: hop-by-hop CoAP deployment protected by OSCORE\@. Response caching and request aggregations are confined to request-response pairs only.
\item Deterministic OSCORE Proxy: hop-by-hop CoAP deployment with deterministic requests and OSCORE protection. Caching and aggregation work across multiple request-response pairs from varying endpoints.
\item NDN\@: hop-by-hop NDN deployment with caching and aggregation working for multiple requests and endpoints.
\end{enumerate}
\begin{table}[h]
  \centering
  \begin{tabular*}{\columnwidth}{lccc}
    \toprule
    \multirow{2}{*}{Protocol} & \multirow{2}{*}{Caching} & \multirow{2}{1.25cm}{\centering Request Aggregation} & \multirow{2}{1.1cm}{\centering Response Fan-out}\\
    &&&\\
    \midrule
    {OSCORE} & --- & --- & ---\\[0.35em]
    {OSCORE Proxy} & {single party} & {only retransmissions} & ---\\[0.35em]
    \multirow{2}{2cm}{Deterministic OSCORE Proxy} & \multirow{2}{*}{multiple parties} & \multirow{2}{*}{multiple parties} & \cmark\\[0.1cm]
    &&&\\
    {NDN} & {multiple parties} & {multiple parties} & \cmark\\[0.25em]
    \bottomrule\\
  \end{tabular*}
  \caption{Multiparty characteristics of the selected deployments.}\label{tab:protocol-settings}
\end{table}

We configure three frame retransmissions with an exponential backoff in the range of milliseconds in case of missing acknowledgments on the link-layer.
Respectively, we enable three request retransmissions for CoAP and NDN using a message timeout of two seconds.
Request buffers are sufficiently dimensioned to not reject retransmitting requests due to unavailable buffer space.
Cache implementations for CoAP and NDN are equivalently scaled to hold 40 responses, which is an adequate number for our setup to not replace up-to-date cache entries required by delayed retransmissions.

\subsubsection{Security Configuration}
The standard OSCORE operation provides peer authentication with a strong message binding between single requests and responses.
To preserve the peer authenticating characteristic of OSCORE in a group-key environment, deterministic OSCORE leverages digital signatures in responses using the Edwards-curve Digital Signature Algorithm (EdDSA) that is carried inside the response payload.
This ensures that members can access and read encrypted messages, but the digital signature prevents forgery attempts by group members.
Software based signature derivations can occupy the processing unit by more than a second, which renders a realistic content retrieval scenario with multiple parties and prominent traffic patterns unusable.
Since the hardware platform in the testbed does not provide hardware-assisted crypto operations, we replicate the signature derivation time of common cryptoprocessors and set it to a constant-time delay of 20~ms~\cite{kblsw-pscli-21} in our server firmware.

\subsection{Comparative Evaluation}

\subsubsection{Content Retrieval Time}
In our first evaluation, we gauge the time from initiating a request on a client to the arrival of the requested content on the application.
This time measurement does not only include pure message round-trips, but additionally scales with packet loss and retransmission events.
\figurename~\ref{fig:completiontime-cdf} summarizes the distributions of content retrieval times for our protocol ensemble and clients$_{1,5,9}$ from \figurename~\ref{fig:topology}.

We observe that all deployments are in agreement with the retransmission behavior.
Successful message deliveries without corrective actions finish in the sub-second range, while retrieval times multiply by the configured retransmission timeout into the seconds range on packet loss.
This leads to a stair case pattern every two seconds for all distributions.
CDF maxima for each protocol marks the overall success rate.

Client$_1$ shows an overall positive success rate for all protocol expressions: more than 96\% of requests succeed for the end-to-end OSCORE deployment, while the hop-by-hop variants increase up to 100\%.
Since client$_1$ retrievals are not affected by cross traffic from other clients, these affirmative results are expected.
The distributions progressively degrade for each client if we move towards client$_9$ due to the increased traffic load that manifests on the shared path between $f_1$ and $f_7$.
Three aspects become visible when we observe the completion time distributions for client$_9$.
First, the multiparty-unaware deployments display reduced success rates of up to $\approx$40\%.
OSCORE Proxy minimally improves on the standard OSCORE setup because of cache hits, which are however confined to request retransmissions from a single client.
NDN and the deterministic OSCORE Proxy deployment open up this restriction and allow the utilization of cached responses by all clients.
This leads to steady success rates across all client nodes independently from the induced traffic load.
Second, the latency in the sub-second range improves drastically from around 100~ms down to $\approx$25~ms for the multiparty-aware deployment variants due to the availability of prepopulated caches.
Third, the latency for request retransmissions in the seconds range equally reduces for NDN and the deterministic OSCORE Proxy variant as displayed by the steep stair cases.
This results from request aggregations on hops that experience loss events and have in turn already scheduled retransmission events.
Returning responses then fan out to all interested clients prematurely well before local retransmission timeouts.

\subsubsection{Cache Utilization and Request Aggregation}
\begin{figure}
  \centering
  \includegraphics{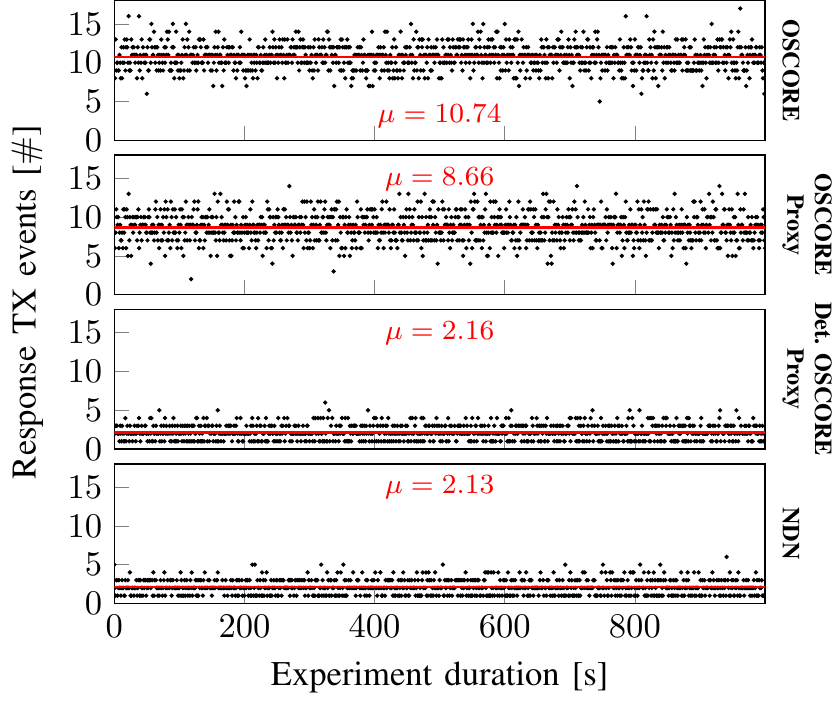}%
  \caption{Quantity of outgoing responses measured at the server node.}%
  \label{fig:server-response-tx-scatter}
\end{figure}

In our next comparison, we measure the server load for our protocol selection to quantify the effects of cache utilization and request suppression.
\figurename~\ref{fig:server-response-tx-scatter} presents the response transmission rate on the server node for the duration of the experiment.

\begin{table*}
  \centering
  \begin{tabular}{lccccccccc}
    \toprule
    \multicolumn{1}{c}{\multirow{2}[3]{*}{Operations}} & \multicolumn{2}{c}{OSCORE} & \multicolumn{2}{c}{OSCORE Proxy} & \multicolumn{2}{c}{Det.\ OSCORE Proxy} & \multicolumn{2}{c}{NDN}\\
    \cmidrule(lr){2-3} \cmidrule(lr){4-5} \cmidrule(lr){6-7} \cmidrule(lr){8-9}
    & Client & Server & Client & Server & Client & Server & Client & Server\\
    \midrule
    Authenticated Encryption & 2.93 (\textcolor{red!47!black}{$\uparrow$47\%}) & 4.63 (\textcolor{red!100!black}{$\uparrow$132\%}) & 2.66 (\textcolor{red!33!black}{$\uparrow$33\%}) & 3.21 (\textcolor{red!61!black}{$\uparrow$61\%}) & 2.08 (\textcolor{red!4!black}{$\uparrow$4\%}) & 0.49 (\textcolor{red!11!black}{$\uparrow$11\%}) & 1.00 (\textcolor{red!0!black}{$\checkmark$}) & 0.23 (\textcolor{red!5!black}{$\uparrow$5\%})\\[0.3em]
    Signature Creation / Verification &  --- & --- & --- & --- & 1.00 (\textcolor{red!0!black}{$\checkmark$}) & 0.24 (\textcolor{red!9!black}{$\uparrow$9\%}) & --- & ---\\[0.3em]
    Message Authentication Code & --- & --- & --- & --- & 3.24 (\textcolor{red!8!black}{$\uparrow$8\%}) & 0.73 (\textcolor{red!9!black}{$\uparrow$9\%}) & 1.00 (\textcolor{red!0!black}{$\checkmark$}) & 0.23 (\textcolor{red!5!black}{$\uparrow$5\%})\\[0.3em]
    \bottomrule\\
  \end{tabular}
  \caption{Mean cryptographic operations per successful content retrieval. In parenthesis, the relative overhead added by packet loss.}\label{tab:cryptoops}
\end{table*}

In accordance with the number of clients and the configured scheduling interval for requests, the optimum rate on the server is 9~packets/s in deployments without cross-client caching.
We observe an approximation of this expected value for the plain OSCORE and OSCORE Proxy setups.
In the first case, the average transmission rate is above ten, which is an indication for request retransmissions, and in the latter case, the rate averages to slightly below the optimum value.
This comes as no surprise, since retransmissions of the same origin can benefit from cache hits in the network.
Requests that never arrive at the server result in a slightly reduced average.

The optimum average rate scales down to 2 packets/s when caching and request aggregation is enabled.
When precisely scheduled, all requests in the subtree containing clients$_{2\text{--}9}$ collapse into a single request that reaches the server and the additional request arrives from client$_{1}$.
Deterministic OSCORE Proxy as well as NDN show in \figurename~\ref{fig:server-response-tx-scatter} an average transmission rate that is comparative to the theoretical optimum.

\subsubsection{Security Effort}

To put computational effort associated with the different protocols into perspective,
we compare the number of performed and necessary cryptographic operations.
Numbers in \tablename~\ref{tab:cryptoops} express operations per successful completion of a request by a single client.
The ideal conditions against which we benchmark are derived from the analysis of the configured topology in \figurename~\ref{fig:topology} assuming no packet loss.

For the OSCORE deployment, the optimal number of AEAD computations is two for clients and the server.
Clients perform an AEAD operation when they generate a request and when they receive a response.
Conversely, servers operate similarly for receiving a request and generating a response.
We observe in \tablename~\ref{tab:cryptoops} that on average clients perform 47\% more AEAD operations when compared against successfully received responses.
This is a result of packet loss, where retransmissions were not able to recover lost response messages.
Note that request retransmissions do not add to the overhead, because they already reside in the retransmission buffer with security related information populated.
The server side shows a computational effort that increases by 132\%.
Arriving request retransmissions lead to multiple generations of a response.

The OSCORE Proxy deployment behaves equally with respect to the necessary cryptographic operations.
Due to the caching and aggregation features that only span a single request-response pair, message losses are reduced.
Effectively, this decreases the amount of unnecessary AEAD operations for clients and the server when compared to OSCORE\@.

In deterministic OSCORE Proxy, the averaged AEAD operations further decrease due to the multiparty-aware protocol operation.
The high success rates on clients yield a nominal AEAD overhead of only 4\%.
Given our topology in \figurename~\ref{fig:topology} and the configured request scheduling, only two of nine requests reach the server per second in ideal conditions.
Requests arriving from the upper subtree collapse into a single request during forwarding and another single request arrives from client$_1$.
The amortized ideal number of AEAD operations on the server is thus $2 \cdot \frac{2}{9} = 0.44$.
In addition to AEAD operations, this deployment requires the creation and verification of digital signatures in responses.
Clients have an optimal value of one signature verification for each successfully received response.
The server creates $\frac{2}{9} = 0.22$ signatures per request round for all clients.
The very low number of retransmissions that actually arrive at the server produce an extra effort of 9\%.
This deployment further uses three HMAC operations per request and response.
Clients demonstrate analogously a small overhead and the server equally shows an added effort of 8\% from the optimal value $3 \cdot \frac{2}{9} = 0.67$.

The NDN deployment presents comparable overheads.
The ideal number of operations differ largely, though:
In this setup, only responses are signed and carry AEAD encrypted content.

For a useful comparison of OSCORE with deterministic OSCORE,
the added asymmetric operations need to be considered with their time or power consumption;
the former outweighing the latter by a factor of over $100$ in both aspects~\cite{kblsw-pscli-21}.
Taken on their own, the energy savings of caching fail to justify the use of deterministic OSCORE with this chosen topology.
However, considering the combined energy use of cryptographic operations and radio transmissions,
where the consumption of a frame transmit plus receive operation (about $10^{-3} J$)
is larger than that of even an asymmetric cryptographic operation (about $0.5 \times 10^{-3} J$),
it is easily seen that the reduced link utilization of deterministic OSCORE amortizes the increased energy cost of the asymmetric cryptography.

\begin{figure}
  \centering
  \includegraphics{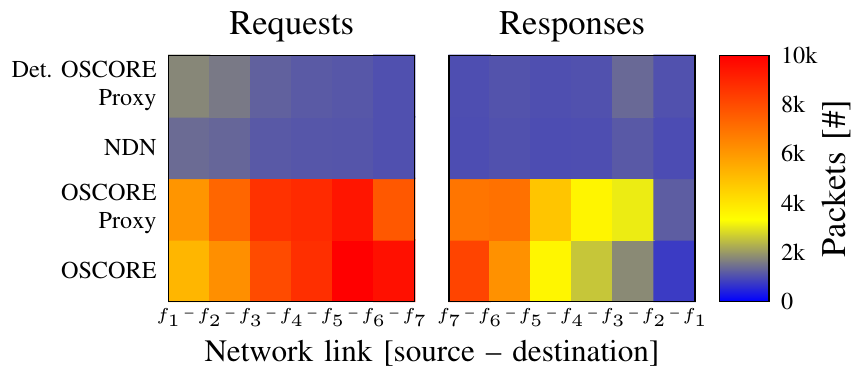}%
  \caption{Requests and responses recorded along the shared path $f_{1\text{--}7}$.}%
  \label{fig:link-stress-matrix}
\end{figure}

\subsubsection{Link Stress}
Our protocol collection shows nuanced variations in the amounts of packet transmissions throughout the topology.
\figurename~\ref{fig:link-stress-matrix} illustrates absolute packet transmissions for the path between forwarder nodes $f_{1\text{--}7}$.
In general and given the intersections of our constructed tree topology in \figurename~\ref{fig:topology}, the amount of request transmissions increases gradually for each hop that is closer to the server.
On the other hand, the amount of transmitted responses between forwarders naturally decreases when moving away from the source.

Each client triggeres 1000 distinct instruction requests, \ie every forwarder $f$ adds another 1000 packets to the upstream path when there is no packet loss in addition to the packets received downstream.
For the OSCORE and OSCORE Proxy configurations, we observe the anticipated continuous increase in requests.
OSCORE Proxy shows slightly more requests per link, because retransmissions take place hop-by-hop, while there is a much higher chance of losing request retransmissions along the path for the end-to-end OSCORE\@.
This characteristic yields insignificantly less link stress for OSCORE, but does not promote success rates as we have observed in earlier evaluations.
Downstream transmissions show similarly expected results:
Unnecessarily large quantities of returning responses appear close to the server and this number declines very fast with each hop for OSCORE due to packet loss.
OSCORE Proxy smooths out the steady decrease in responses with the use of response caching for retransmissions of the original request.
The multiparty-aware protocols operate comparably with small differences resulting from larger packet sizes for the deterministic OSCORE deployment.
The pluralistic cache utilization and request aggregation features allow both variations to reduce the packet transmissions on each forwarder hop to approximately 1000--2000 packets.


\section{Conclusions and Outlook}\label{sec:conclusion}
Information-centric content replication has repeatedly proven beneficial in low-power  networks.
Hop-by-hop forwarding, in-network caching, and hop-wise retransmissions are key promoters of reliably delivering packets in lossy wireless regimes. This work is part of an ongoing effort to develop an information-centric Web of Things (WoT) that is built on CoAP and OSCORE\@. In this paper, we designed protocol configurations and extensions that carry one-to-many CoAP data flows with OSCORE content object security. 

In detail, we first explored missing multiparty protocol features, then identified protocol entry-points for our extensions, and finally laid out a blueprint for integrating  \one multi-destination forwarding with response de-duplication, \two request aggregations with response fan-outs, and \three a pluralistic cache utilization.

Using full protocol implementations on RIOT OS and a real-world testbed, we comparatively evaluated our multiparty communication model against CoAP OSCORE (w/ and w/o Proxy) and NDN. We found significant improvements in network utilization and a much reduced link stress. On the overall, we could show that multiparty content dissemination works equally efficient with our CoAP-based WoT solution and NDN\@. The results confirm that information-centric principles can be built into the CoAP ecosystem without sacrificing interoperability nor performance.  

The following items are on our future research agenda: 

\begin{enumerate}
\item Extensions of the CoRE Resource Directory~\cite{draft-ietf-core-resource-directory} to inquire groups and support multi-destination routing.
\item Translation of CoAP unicast requests to IP multicast at CoAP proxies~\cite{draft-tiloca-core-groupcomm-proxy}, which  integrates the IP multicast domain with regular CoAP unicast requests.  
\end{enumerate}

Integrating these additional protocol elements will further complete the architecture of a data-centric Web of Things and make it versatile and useful to foster an open, standard-compliant Internet of Things.


\paragraph{Acknowledgments}
This work was supported in part by the German Federal Ministry for Education and Research (BMBF) within the projects \textit{I3 -- Information Centric Networking for the Industrial Internet} and the Hamburg \textit{ahoi.digital} initiative with \textit{SANE}.
The \texttt{libOSCORE} library was made with financial support from Ericsson AB\@.


\balance%
\bibliographystyle{IEEEtran}
\bibliography{}

\end{document}